\documentclass[prb,superscriptaddress,floatfix,twocolumn]{revtex4-1}
\usepackage{graphicx}
\usepackage{color}

\def\lsco{La$_{2-x}$Sr$_x$CuO$_4$}
\def\lbco{La$_{2-x}$Ba$_x$CuO$_4$}

\def\ybco{YBa$_2$Cu$_3$O$_{6+x}$}

\def\bscco{Bi$_2$Sr$_2$CaCu$_2$O$_{8+\delta}$}

\def\tbkt{$T_{\rm BKT}$}

\begin{document}

\title{Evolution of superconducting correlations within magnetic-field-decoupled\\ CuO$_2$ layers of La$_{1.905}$Ba$_{0.095}$CuO$_4$}

\author{Z. Stegen}
\affiliation{National High Magnetic Field Laboratory, Florida State University, Tallahassee, FL 32310}
\author{Su Jung Han}
\affiliation{Condensed Matter Physics \&\ Materials Science Department, Brookhaven National Laboratory, Upton, NY 11973-5000}
\affiliation{Department of Materials Science and Engineering,
Stony Brook University, Stony Brook, New York 11794, USA}
\author{Jie Wu}
\affiliation{National High Magnetic Field Laboratory, Florida State University, Tallahassee, FL 32310}
\affiliation{Condensed Matter Physics \&\ Materials Science Department, Brookhaven National Laboratory, Upton, NY 11973-5000}
\author{A. K. Pramanik}
\affiliation{Institute for Solid State Research, IFW Dresden, D-01171 Dresden, Germany}
\author{M. H\"ucker}
\author{Genda Gu}
\author{Qiang Li}
\affiliation{Condensed Matter Physics \&\ Materials Science Department, Brookhaven National Laboratory, Upton, NY 11973-5000}
\author{J. H. Park}
\author{G. S. Boebinger}
\affiliation{National High Magnetic Field Laboratory, Florida State University, Tallahassee, FL 32310}
\author{J. M. Tranquada}
\affiliation{Condensed Matter Physics \&\ Materials Science Department, Brookhaven National Laboratory, Upton, NY 11973-5000}
\date{\today}
\begin{abstract}
We explore the evolution of superconductivity in \lbco\ with $x=0.095$ in magnetic fields of up to 35~T applied perpendicular to the CuO$_2$ planes.  Previous work on this material has shown that perpendicular fields enhance both charge and spin stripe order within the planes.  We present measurements of the resistivity parallel and perpendicular to the planes, as well as the Hall effect.  Measurements of magnetic susceptibility for fields of up to 15 T applied both parallel and perpendicular to the planes provide complementary measures of the superconductivity.  We show that  fields sufficient to destroy pair tunneling between the planes do not disrupt the superconducting correlations within the planes.  In fact, we observe an onset of large amplitude but phase disordered superconductivity within the planes at approximately 30~K that is remarkably insensitive to field.  With further cooling, we observe a phase-transition-like drop in the in-plane resistivity to an apparent state of superconductivity, despite the lack of phase coherence between the layers.  These observations raise interesting questions concerning the identification of the upper critical field, where pairing is destroyed, in underdoped cuprates.  
\end{abstract}
\maketitle

\section{Introduction}

Since Emery and Kivelson\cite{emer95a} first suggested that phase fluctuations might limit the  transition temperature $T_c$ of superconductors with low superfluid density, there has been considerable research aimed at determining where and how strong superconducting correlations turn on in cuprates.  This issue applies both to the onset of superconductivity on cooling and to the loss of superconductivity in an increasing magnetic field.  The topic has both theoretical and practical relevance.  On the theoretical side, there continue to be questions regarding the extent to which the onset of pairing correlations might be connected with the pseudogap phenomena in underdoped cuprates.  On the practical side, if there are regions of temperature and field where only phase fluctuations limit superconducting order, one might hope to find ways to enhance phase order so as to extend the useful range of superconducting order.

Experimental evidence for superconducting fluctuations at temperatures far above $T_c$ have been provided by Nernst effect and torque magnetometry measurements on a variety of cuprates by Ong and coworkers.\cite{xu00,wang01,wang02,wang06,li05,li07b,li10}  This work motivated theoretical suggestions\cite{lee06,ande08,tesa08} of a possible phase-disordered 2D superconducting state (2D vortex liquid) that might exist above $T_c$.    In contrast, recent studies of superconducting contributions to magnetoresistance\cite{rull11} and low-frequency optical conductivity,\cite{bilb11a,bilb11b} as well as further torque magnetometry studies,\cite{mosq11,yu12} indicate that strong superconducting correlations are found only within a relatively narrow range ($\sim 10$~K) above $T_c$; the response at higher temperatures is quite weak relative to expectations for a 2D vortex-liquid state.\cite{asla68}

There is general agreement that pairing interactions within the CuO$_2$ layers are responsible for the development of superconducting correlations in the cuprates.  Josephson coupling between the layers leads to the onset of  three-dimensional (3D) superconductivity as soon as the correlation length for superconducting order within the layers becomes sizable.\cite{lawr71,klei92}  The recent observations\cite{rull11,bilb11a,bilb11b,mosq11,yu12} that strong superconducting correlations appear only in a regime that is reasonably close to $T_c$ are consistent with the expectation that 3D order should appear as soon as 2D superconducting correlations become substantial.\cite{lawr71}  In fact, in a previous study of \lbco\ (LBCO) with $x=0.095$, evidence was found indicating that superconducting correlations between layers start to develop locally before superconducting correlations diverge within the layers.\cite{wen12a} Nevertheless, there remains a question as to whether one might be able to observe the 2D vortex-liquid state by suppressing the interlayer Josephson coupling with a magnetic field applied perpendicular to the layers.

Another question concerns the evolution of superconducting correlations as order is suppressed by a strong magnetic field.  For a type-2 superconductor, the initial onset of finite resistivity corresponds to the flow of vortices; destruction of Cooper pairs should occur at a higher field, conventionally labeled $H_{c2}$.  Given the large magnitude of the superconducting gap in underdoped cuprates, one might expect $H_{c2}$ to be much larger than the field at which resistivity appears; however, a variety of recent transport measurements on \ybco\ have been interpreted in terms of a rather low $H_{c2}$, especially in the vicinity of a hole concentration of 1/8.\cite{chan12b}   This is the same regime where quantum oscillations have been observed in the high-field state.\cite{tail09,seba12}  The quantum oscillations are a response of normal quasiparticles; however, such a response can occur in the mixed state of a superconductor as well as in the normal state.\cite{norm10,bane13}  Indeed, a specific-heat study indicates that the quantum oscillations do occur within the superconducting mixed state.\cite{rigg11}  To explain the measured cyclotron frequency, reconstruction of the Fermi surface by competing order has been invoked.\cite{tail09,seba12}  Several recent experiments have provided direct evidence for charge order that is enhanced when superconductivity is depressed by a strong magnetic field.\cite{wu11,ghir12,chan12a}  The relationship between the superconductivity, charge order, and normal quasiparticles remains a hot topic of debate.

In this article, we present a study of the superconductor La$_{1.905}$Ba$_{0.095}$CuO$_4$ ($T_c=32$~K) in strong magnetic fields $H_\bot$ applied perpendicular to the CuO$_2$ planes.  In previous work, it has been shown that the weak charge stripe order present in zero field\cite{huck11} is enhanced by $H_\bot$.\cite{wen12,huck13}  Here we demonstrate that a strong enough $H_\bot$ can completely destroy the phase coherence between neighboring layers without destroying the superconducting correlations within the layers.  Evidence for the layer decoupling is obtained from measurements of the resistivity perpendicular to the layers, $\rho_\bot$, while evidence for the survival of the superconductivity is provided by measurements of magnetic susceptibility, Hall effect, and resistivity parallel to the layers, $\rho_\|$.  We find that the onset of strong superconducting correlations within the decoupled layers occurs at approximately 30~K, with little variation due to $H_\bot$ up to our maximum of 35~T.  We label this a layered vortex liquid (LVL) state; it is essentially a 2D vortex liquid state, but there could be electromagnetic interactions between the layers associated with the vortices, resulting in 3D correlations.\cite{rama09}

\begin{figure}[t]
\centerline{\includegraphics[width=3.5in]{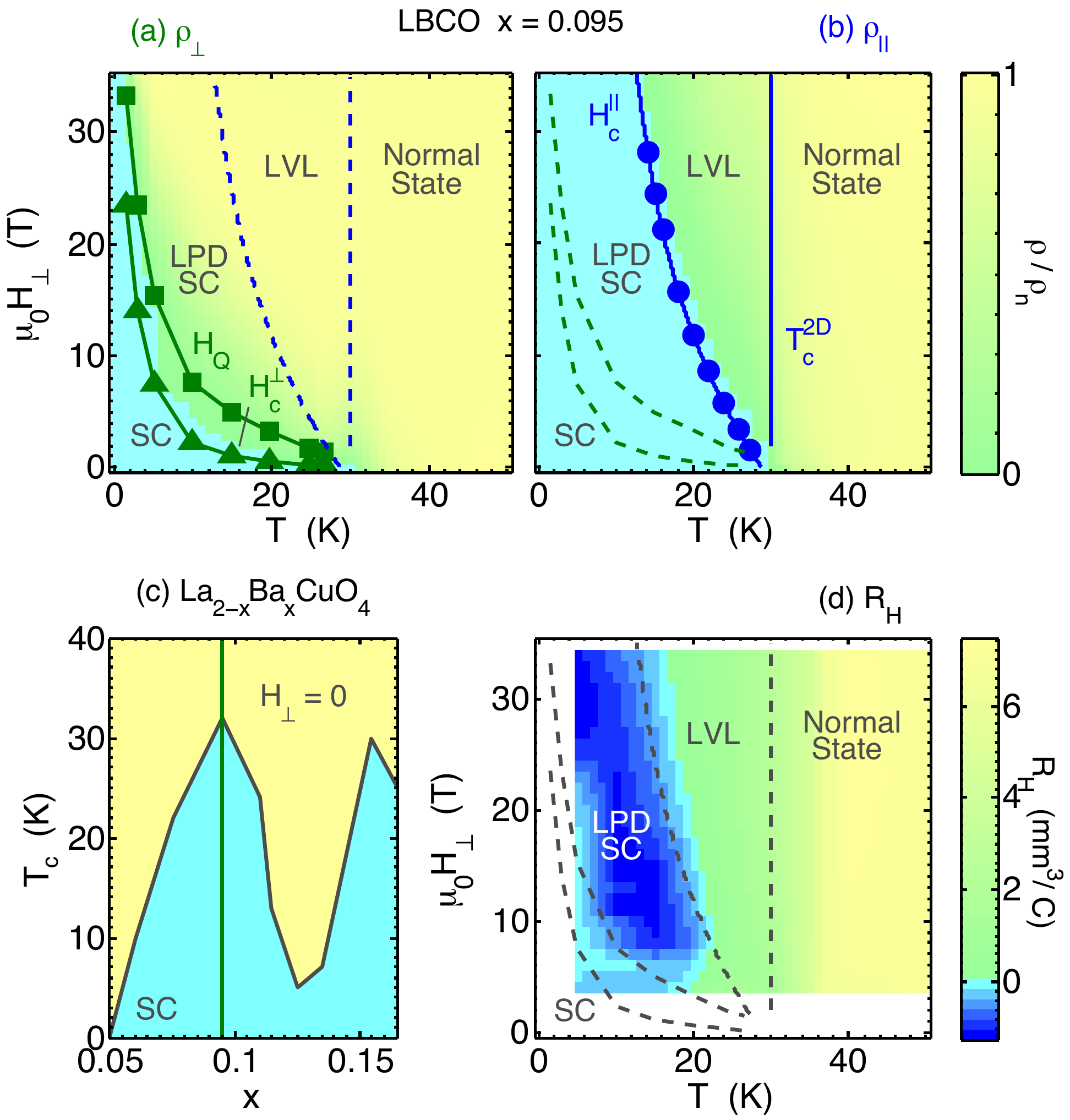}}
\caption{(color online) Phase diagrams in $H_\bot$-$T$ space obtained from measurements of (a) $\rho_\bot$ and (b) $\rho_\|$.  In (a), triangles indicate the onset of finite $\rho_\bot$ at $H_c^\bot$; squares denote $H_Q$, corresponding to the interlayer phase-decoupling crossover.  In (b),  circles indicate onset of finite $\rho_\|$ at $H_c^\|$; vertical solid line corresponds to $T_c^{\rm 2D}$, the crossover from the layered vortex liquid (LVL) phase to the normal state.  In both (a) and (b), the shaded contours correspond to the resistivity normalized to an extrapolation of the normal-state behavior obtained at the maximum field.  (c) Doping dependence of $T_c$ in \lbco, from Ref.~\onlinecite{huck11}; vertical line denotes present sample.  (d) Hall coefficient, $R_{\rm H}$, in $H_\bot$-$T$ space.  For panels (a)-(c), the zero-resistance state ($\rho/\rho_n<10^{-3}$) corresponds to the regions in cyan (online).  Note that the region of negative Hall constant [dark blue in (d)]
corresponds to the regime of layered phase-decoupled superconductivity (LPD SC) with finite $\rho_\bot$ and negligible $\rho_\|$.}
\label{fg:phase_diag} 
\end{figure}

Within the LVL state, $\rho_\|$ has a finite magnitude consistent with that expected for a 2D superconductor without phase order.\cite{asla68}  On cooling in fixed field, $\rho_\|$ decreases in a fashion suggesting critical behavior similar to that predicted\cite{halp79} for a 2D superconductor on the approach to the phase-ordering transition of Berezinkii\cite{bere71} and Kosterlitz and Thouless\cite{kost73} (BKT).  (We note that the theory applies only to the case of zero field.)  Following the variation of $\rho_\|$ with $H_\bot$ at fixed temperature, we observe behavior suggesting a transition to a state with negligible $\rho_\|$ despite an absence of phase coherence between the layers.  We label this state a layered, phase-decoupled superconductor (LPD-SC).   Our results are summarized in Fig.~\ref{fg:phase_diag}.

The occurrence of the LPD-SC state (regardless of whether true superconducting order is achieved), as well as the complete decoupling of the layers in the LVL state, indicates a frustration of the interlayer Josephson coupling\cite{wen12} by some mechanism other than thermal vortex fluctuations.\cite{glaz91,kosh96a,kosh96b}  Together with the field-enhanced charge-stripe order,\cite{wen12} there is a clear parallel with behavior reported for LBCO with $x=1/8$, where LVL and LPD-SC states associated with stripe order were observed in zero field.\cite{li07,tran08}  In that case, the frustration of the Josephson coupling has been explained in terms of a proposed pair-density-wave (PDW) superconductor.\cite{hime02,berg07,berg09b}  The similar phenomenology suggests that field-induced PDW order could be relevant to the $x=0.095$ sample.

The rest of this paper is organized as follows.  The experimental methods are described in the following section.  In Sec.~\ref{sc:normal}, we present the resistivity data and analyze the normal-state magnetoresistance.  Evidence for the onset of strong superconducting correlations within the CuO$_2$ layers from Hall effect and magnetic susceptibility measurements is presented in Secs.~\ref{sc:hall} and \ref{sc:chi}, respectively.  Analysis of $\rho_\bot(H_\bot)$ and the decoupling of the layers is described in Sec.~\ref{sc:perp}, while the evidence from $\rho_\|$ measurements for the transition to the LPD-SC state is given in Sec.~\ref{sc:par}.  The paper concludes with a summary and discussion in Sec.~\ref{sc:summ}.

\section{Experimental method}

The crystals, grown by the traveling-solvent floating-zone method,  have been characterized in several previous studies \cite{huck11,wen12,wen12a,home12}.  Most of the present experiments were performed in the 35-T dc magnet at the National High Magnetic Field Laboratory (NHMFL).  The crystals for the measurements of $\rho_\bot$ and $\rho_\|$ are the same as those used in a previous transport study \cite{wen12}, and the contact configurations are described there.  The resistance was measured using an ac resistance bridge with an excitation current of 1 mA.  All measurements were done after field-cooling from above $T_{c0}$, sweeping the field from 35 T to 0 while holding the temperature fixed.  (Note that sweeping the field at fixed temperature minimizes the energy consumption of the magnet compared to sweeping the temperature at fixed field.) 

A third crystal was prepared for measurements of the Hall effect.  The geometry was similar to that of the crystal for the $\rho_\|$ measurement, except that the voltage contacts were on opposite edges of the crystal in order to measure the Hall voltage $V_{\rm H}$ in the direction transverse to the current flow.  For each measurement, the sample was cooled in zero field from above $T_c$.  Once the temperature was stabilized at the desired value the field was swept from 0 to 35~T, back to 0, down to $-35$~T, and back to 0 again.  In order to eliminate the magnetoresistance contribution due to imperfect alignment of the voltage contacts, the net Hall voltage was calculated as:
\begin{equation}
 V_{\rm H} = [V(+B\uparrow) - V(-B\downarrow)]/2,
\end{equation}
where $V(+B\uparrow)$ corresponds to the up-sweep from 0 to 35~T and $V(-B\downarrow)$ corresponds to the down-sweep from 0 to $-35$~T.  (We checked that the results were the same using the opposite set of field sweeps.)  

The results are expressed in terms of the Hall coefficient $R_{\rm H}$:
\begin{equation}
 R_{\rm H} = {V_{\rm H}d/IB}
\end{equation}
where $d$ is the sample thickness and $I$ is the longitudinal current.  To reduce the noise in the data, the measurements were averaged over windows of 1.75~T in width.  Also, in some cases there were anomalous features at low field, so we present the results just for fields above 3.5~T.

Magnetic susceptibility measurements on a fourth crystal were performed with fields of 7 and 15~T using a vibrating sample magnetometer (VSM) located at the IFW Dresden.  These data have been compared with previous measurements at fields of 1 and 7~T obtained with a SQUID (superconducting quantum interference device) magnetometer at Brookhaven.\cite{huck08}  The temperature dependence of the different measurements at 7 T are in good agreement, but there are small rigid shifts between the data sets.  For presentation, the VSM data have been shifted ($<0.05\times 10^{-4}$ emu/mol) to match the SQUID data at 100 K, where the magnetization is linear in the applied field.

\begin{figure}[t]
\centerline{\includegraphics[width=3.3in]{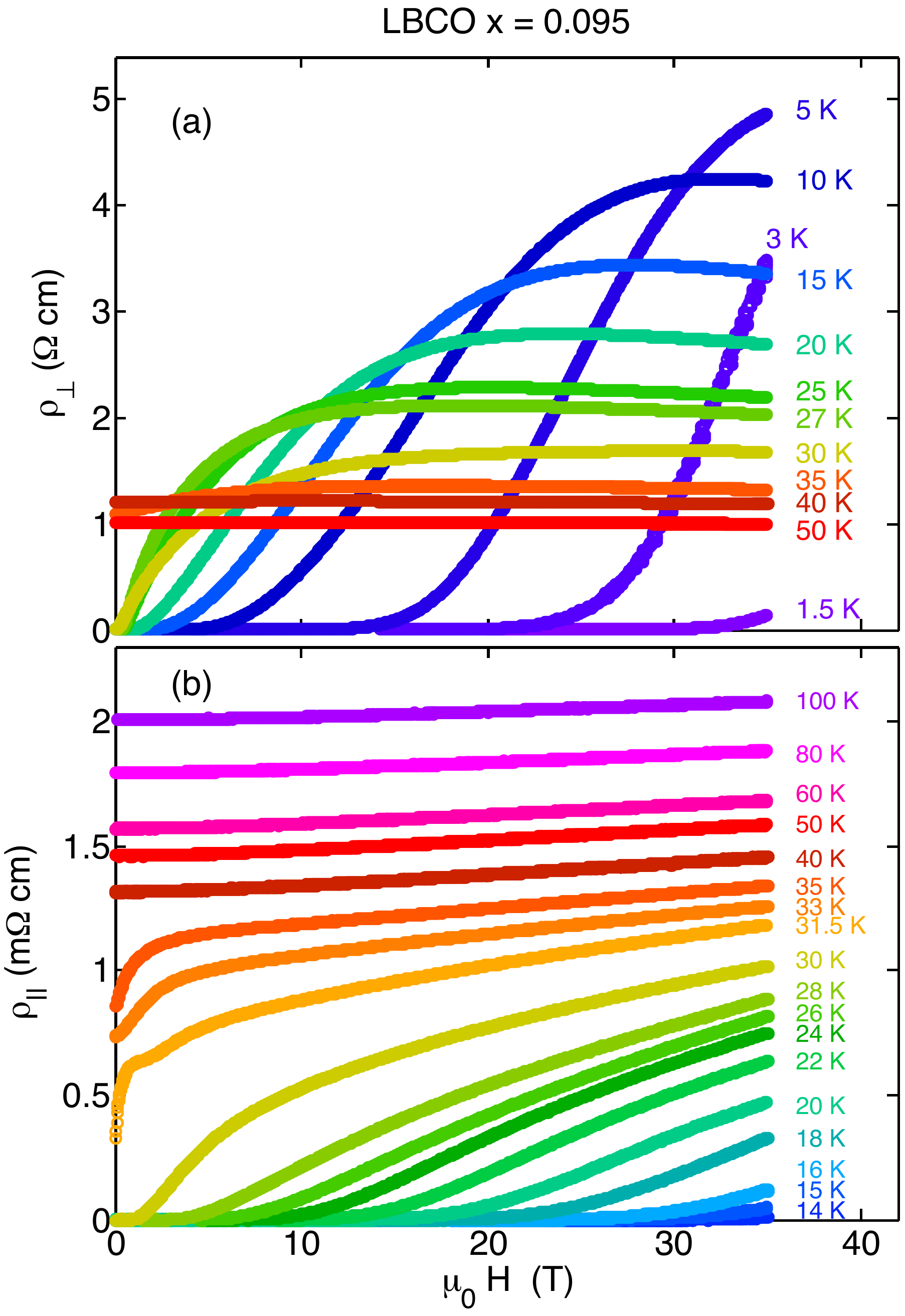}}
\caption{(color online) Measurements of (a) $\rho_\bot$ and (b) $\rho_\|$ as a function of $H_\bot$, obtained at various fixed temperatures as listed in each panel. }
\label{fg:raw} 
\end{figure}

\section{Data and analysis}

\subsection{Normal state and superconducting fluctuations}
\label{sc:normal}

\begin{figure}[t]
\centerline{\includegraphics[width=3.0in]{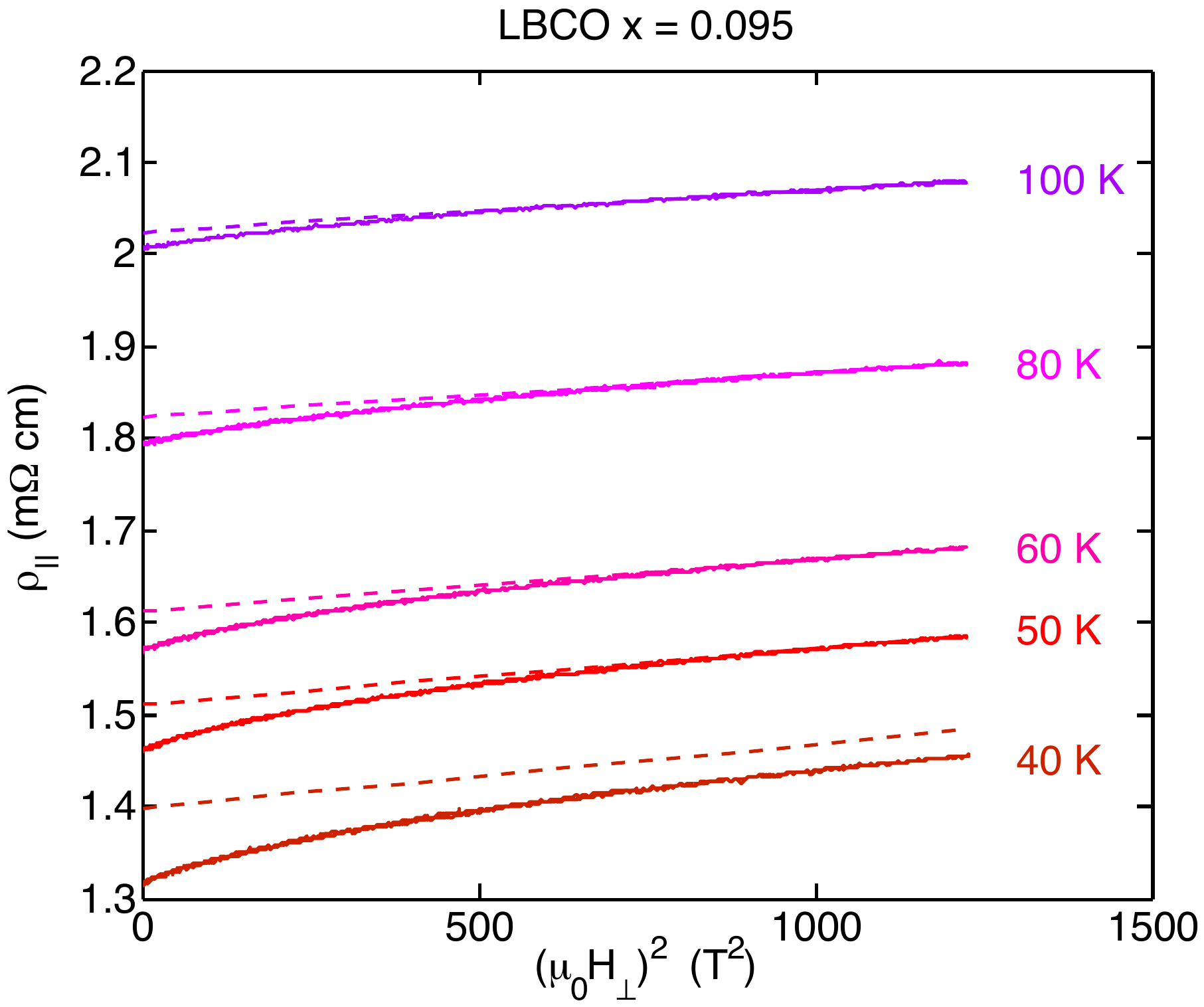}}
\caption{(color online) Plot of $\rho_\|$ (solid lines) as a function of $H_\bot^2$ for $T\gtrsim T_c$, with temperatures listed to the right.  Dashed lines correspond to fits of Eq.~(\ref{eq:lin}) to data for $(\mu_0H_\bot)^2 > 1000~\mbox{\rm T}^2$, except for $T=40$~K, where the value of $\rho_n(H_\bot=0)$ was replaced with the linearly extrapolated value from Fig.~\ref{fg:param}(b).}
\label{fg:rab_hsq} 
\end{figure}

The results for resistivity vs.\ $H_\bot$ obtained for a range of temperatures are shown in Fig.~\ref{fg:raw}.  Let us first consider the data for $T>T_c$.  As discussed by Rullier-Albenque {\it et al.},\cite{rull11} one expects that the in-plane magnetoresistance increases as $H_\bot^2$ in the normal state.\cite{harr95}  Plotting $\rho_\|$ versus $H_\bot^2$ in Fig.~\ref{fg:rab_hsq}, we see that the expected behavior is approached at high fields.  The dashed lines for 50~K and above are fits to the high-field data corresponding to
\begin{equation}
  \rho_{\|,n} = \rho_{\|,n}(H_\bot=0) + a_\rho (\mu_0H_\bot)^2.
  \label{eq:lin}
\end{equation}
The deviations at low field are attributed to superconducting fluctuations.  

\begin{figure}[b]
\centerline{\includegraphics[width=3.0in]{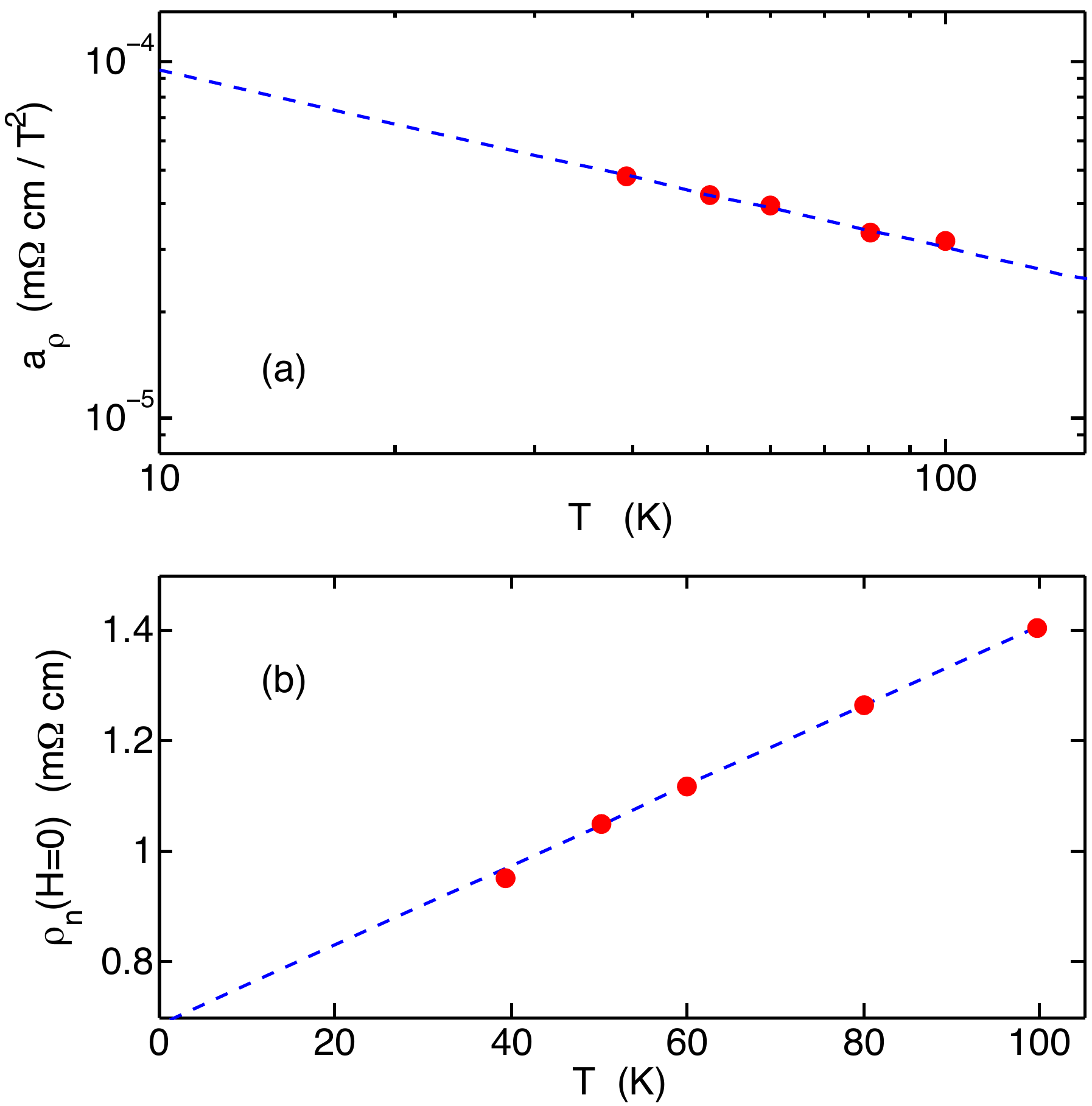}}
\caption{(color online) (a) Coefficient $a_\rho$ vs.\ $T$ obtained from the fits shown in Fig.~\ref{fg:rab_hsq}.  The dashed line has a slope of $-0.5$.  (b)  Plot of the $\rho_{\|,n}(H_\bot=0)$ vs.\ $T$ from the fits in Fig.~\ref{fg:rab_hsq}.  The dashed line is a fit to the points for 50~K $\le T\le 100$~K.}
\label{fg:param} 
\end{figure}

The fitted parameters $a_\rho$ and $\rho_{\|,n}(H=0)$ are plotted in Figs.~\ref{fg:param} (a) and (b), respectively.  The coefficient $a_\rho$ is observed to vary as $T^{-0.5}$.  The quantity $\rho_{\|,n}(H=0)$ varies linearly with temperature for the data from 50 to 100~K.  The downward deviation of $\rho_{\|,n}(H=0)$ at 40~K is correlated with the onset of 3D superconducting fluctuations as demonstrated by Wen {\it et al.}\cite{wen12a}  To approximate the normal-state behavior, we will use the linear extrapolation of $\rho_{\|,n}(H=0)$ from the trend at $T\ge 50$~K.  The extrapolated result at 40 K leads to the dashed line shown in Fig.~\ref{fg:rab_hsq}.  

\begin{figure}[b]
\centerline{\includegraphics[width=3.3in]{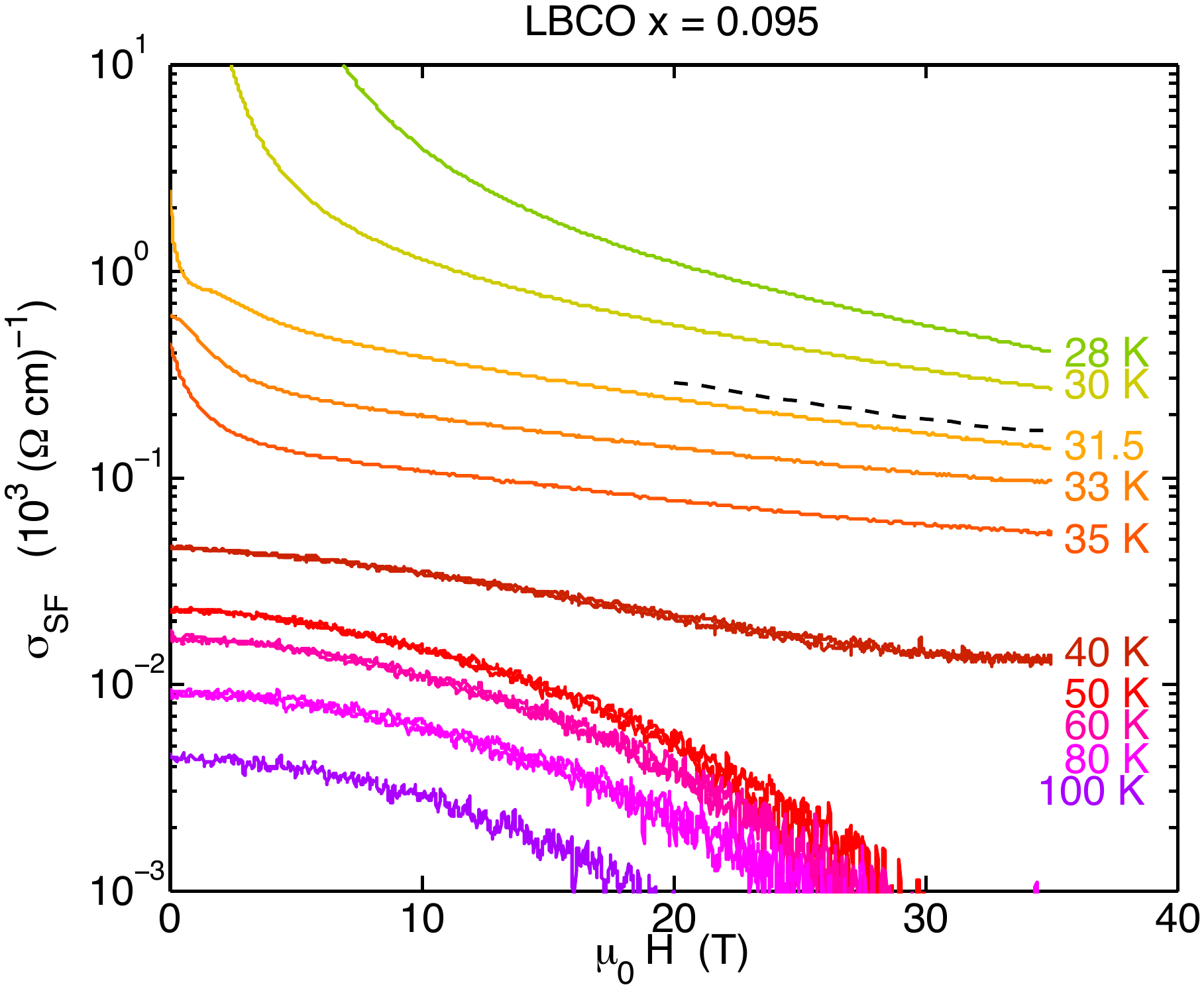}}
\caption{(color online) Conductivity due to superconducting fluctuations determined with Eq.~(\ref{eq:sf}) for $T\gtrsim T_c$.  The dashed line indicates the calculated conductivity from fluctuations in a 2D superconductor according to Eq.~(\ref{eq:AL}) from Ref.~\onlinecite{asla68}, assuming $T=30$~K as discussed in the text. }
\label{fg:SF} 
\end{figure}

Using the extrapolated normal state behavior and assuming a two-fluid model, we can extract the conductivity due to superconducting fluctuations as
\begin{equation}
  \sigma_{\rm SF}(H_\bot,T) = 1/\rho_\| - 1/\rho_{\|,n}.
  \label{eq:sf}
\end{equation}
The results are shown in Fig.~\ref{fg:SF} for a range of temperatures.  At high temperatures, $\sigma_{\rm SF}$ decreases substantially with increasing field.   In contrast, there is a distinct change as one develops 3D superconducting correlations at 40~K and below.  Not only does $\sigma_{\rm SF}(H=0)$ rapidly grow large, but one also observes that the maximum magnetic field is not sufficient to fully suppress $\sigma_{\rm SF}$. 

\begin{figure}[t]
\centerline{\includegraphics[width=3.0in]{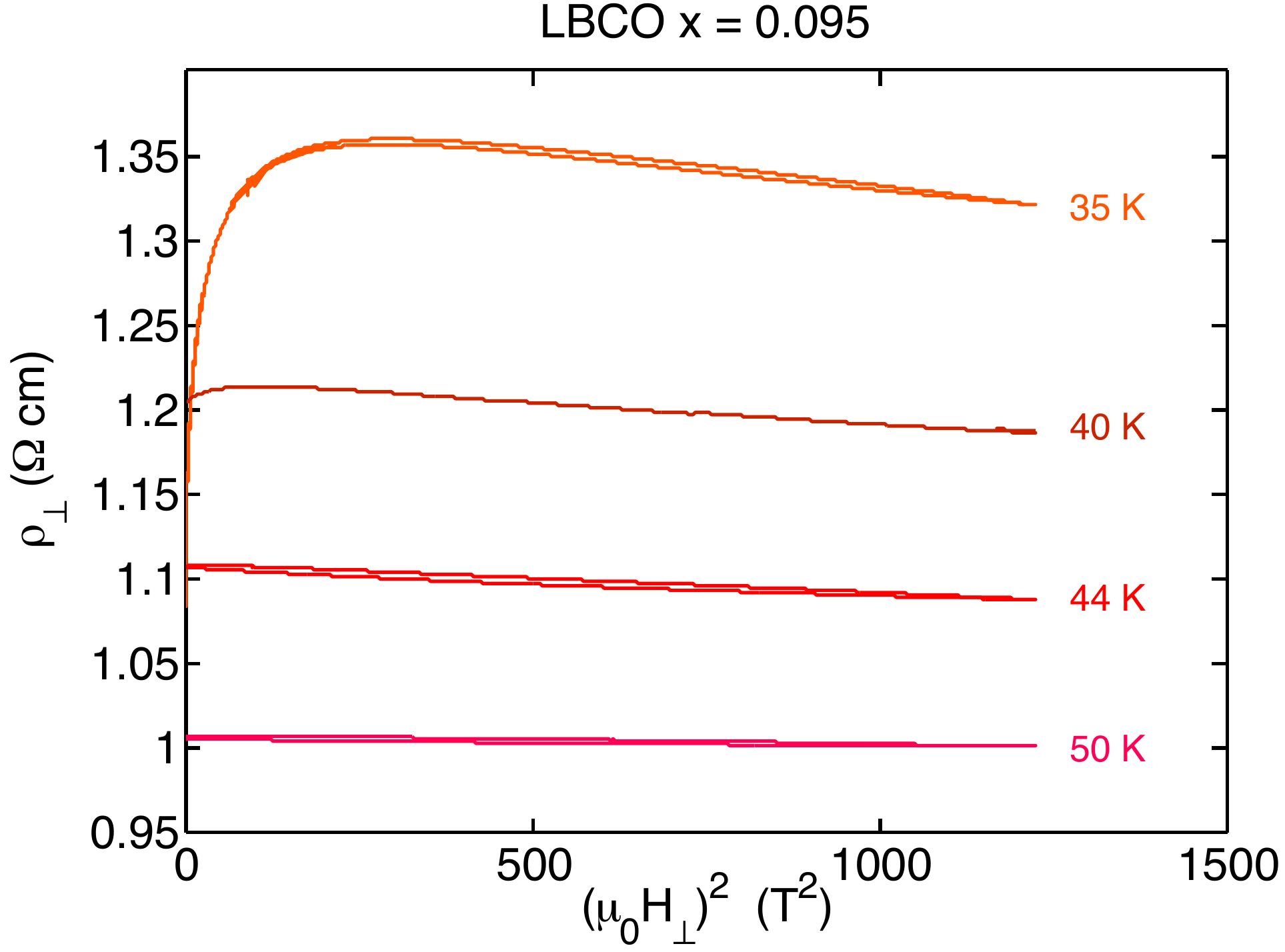}}
\caption{(color online) Plot of $\rho_\bot$ as a function of $H_\bot^2$ for $T\gtrsim T_c$, including sweeps of both increasing and decreasing field. }
\label{fg:rc_hsq} 
\end{figure}

For comparison, we have plotted $\rho_\bot$ versus $H_\bot^2$ in Fig.~\ref{fg:rc_hsq}.  We see that the magnetoresistance is relatively small at 50~K, and a negative magnetoresistance (at high fields) develops on cooling.  For low fields, there is positive magnetoresistance at 40 K and 35 K, indicative of 3D superconducting fluctuations.\cite{wen12a}  The change in sign of the magnetoresistance indicates that the 3D superconducting fluctuations are suppressed at high field; at 35 K, the suppression occurs for $\mu_0H_\bot\gtrsim 20$~T.

Returning to Fig.~\ref{fg:SF}, we see that $\sigma_{\rm SF}$ at 35 K drops significantly on applying a relatively small field, corresponding to suppression of the 3D correlations, but it remains substantial at high field.   A similar pattern is apparent at lower temperatures, as well.  The conductivity due to superconducting fluctuations that survives at high field must occur only within the CuO$_2$ layers.  To evaluate the magnitude of $\sigma_{\rm SF}$, we can compare with the formula obtained by Aslamazov and Larkin\cite{asla68} for a 2D superconductor:
\begin{equation}
  \sigma_{\rm SF} = e^2/16d\hbar \tau,
  \label{eq:AL}
\end{equation}
where $d$ is the thickness of the superconductor and $\tau=(T-T_c)/T_c$.  We take the thickness to be equal to the layer spacing,\cite{lawr71} $s=6.6$~\AA.  The formula was nominally derived for zero field; we will assume that the only impact of the applied field is to reduce $T_c$, as illustrated in Fig.~\ref{fg:phase_diag}(b).  Evaluating the formula for $T=30$~K and $\mu_0H_\bot\ge 20$~T yields the dashed line shown in Fig.~\ref{fg:SF}, which falls about a factor of two below the data curve.  Thus, at 30~K and below, the magnitude of the experimentally-determined $\sigma_{\rm SF}$ at high fields is larger than the prediction for fluctuation pairing in 2D layers.  At higher temperatures, the magnitude of $\sigma_{\rm SF}$ falls off much faster than predicted by Eq.~(\ref{eq:AL}); this is consistent with the conclusion of Rullier-Albenque {\it et al.}\cite{rull11} for superconducting fluctuations in the normal state of \ybco.

\subsection{Hall coefficient, $R_{\rm H}$}
\label{sc:hall}

\begin{figure}[t]
\centerline{\includegraphics[width=3.3in]{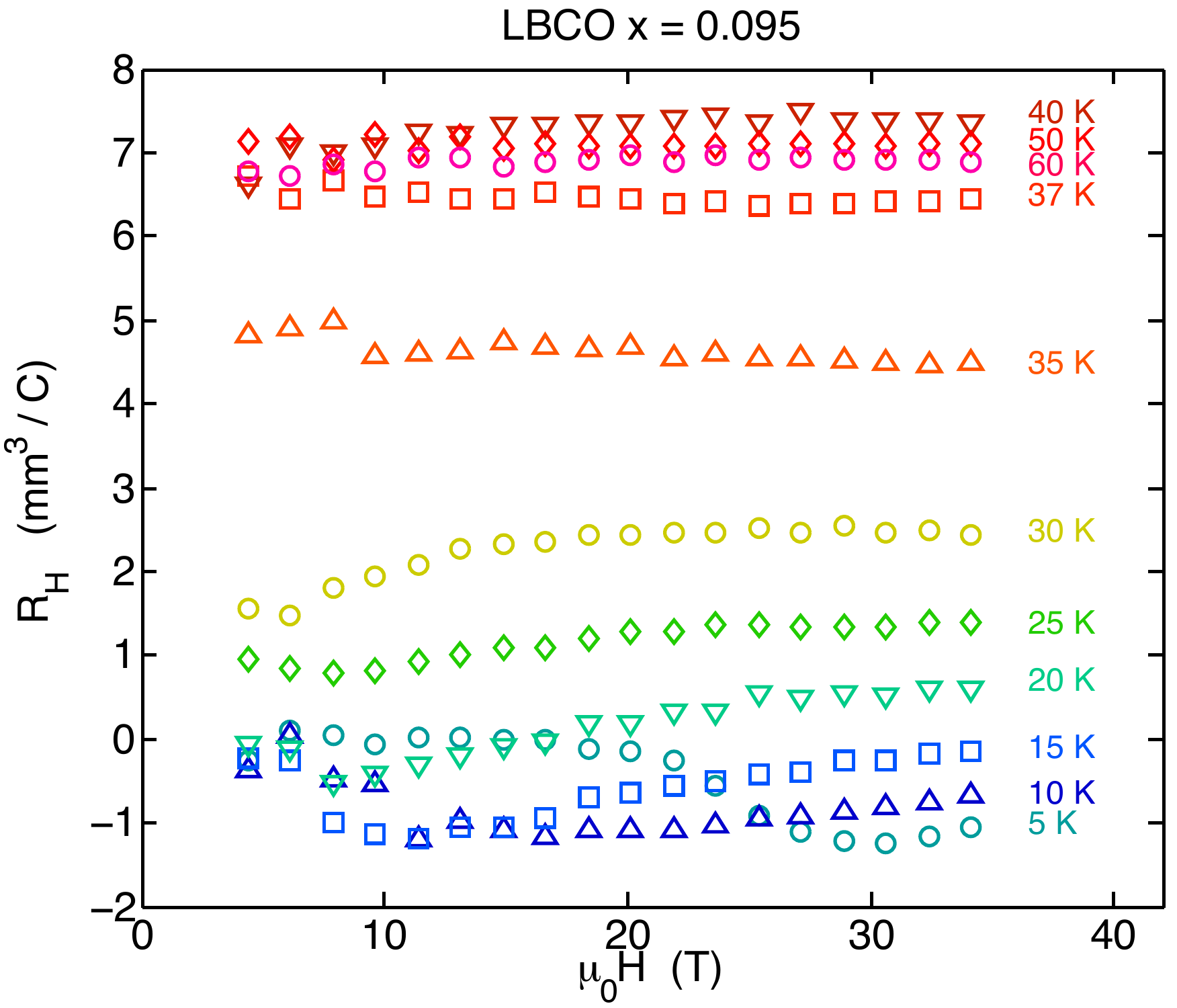}}
\caption{(color online) Data for $R_{\rm H}$  vs.\ $H_\bot$ for various temperatures.  Above 35~K, $R_{\rm H}$ is essentially independent of field; below 25~K, there is notable field dependence, with $R_{\rm H}$ dipping negative.}
\label{fg:Hall} 
\end{figure}

\begin{figure}[b]
\centerline{\includegraphics[width=3.3in]{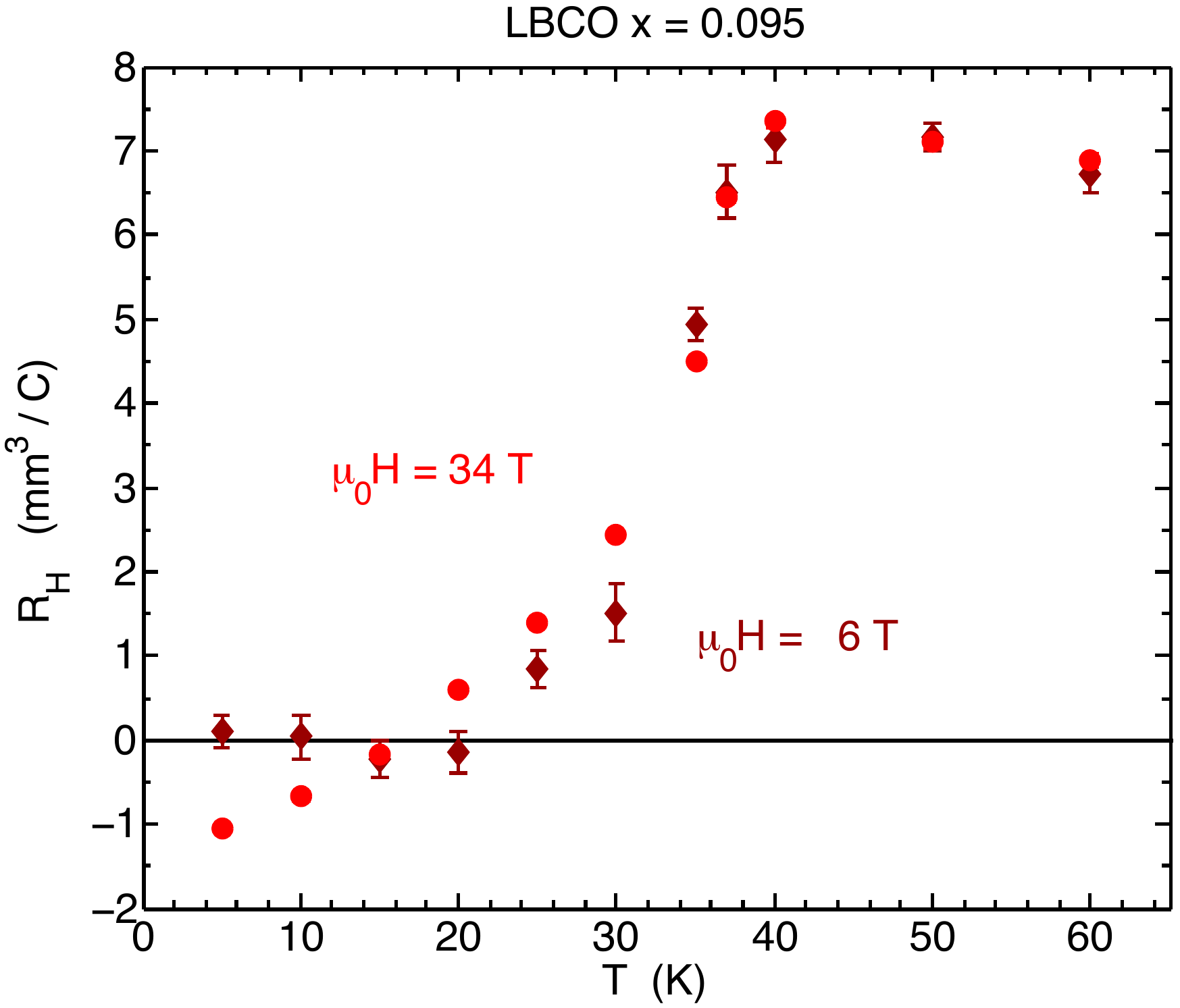}}
\caption{(color online) Data for $R_{\rm H}$  vs.\ $T$ for $\mu_0H_\bot=6$~T and 34~T.  The drop below 40~K correlates with the growth of strong superconducting correlations within the planes. }
\label{fg:Hall_T} 
\end{figure}

Before continuing the analysis of the resistivity data to lower temperatures, let us consider the Hall data.  The full set of measurements is plotted in Fig.~\ref{fg:Hall}.  The field dependence of $R_{\rm H}$ is relatively small compared to the temperature dependence.  To further illustrate this, we compare the temperature dependence of $R_{\rm H}$ for fields of 6 and 34~T in Fig.~\ref{fg:Hall_T}.  In the normal state, $R_{\rm H}$ increases on cooling.  It reaches a maximum near 40~K, below which it rapidly drops in magnitude.  This drop is independent of field.  Very similar behavior has been observed previously for the in-plane thermopower divided by temperature, for fields up to 9~T.\cite{wen12a}  It appears that this drop is due to the rapid growth of in-plane superconducting correlations.   The drop in $R_{\rm H}$ is insensitive to the presence of 3D superconducting correlations, as the same  the initial drop in $R_{\rm H}$ occurs for $\mu_0H_\bot=34$~T, where, as we will see, it must be due to superconducting correlations that are restricted to the CuO$_2$ layers.

We also observe that $R_{\rm H}$ goes negative at low temperature.  Previous studies\cite{adac01,adac11} of LBCO have reported a negative $R_{\rm H}$ below $T_{c0}$ for $x=0.10$ and 0.11, but $R_{\rm H}$ tends toward zero (without going negative) for $x=0.083$ and 0.12.  For our $x=0.095$ sample, the regime of negative $R_{\rm H}$ corresponds to the LPD-SC state, where superconducting order appears within but not between the planes, as indicated in Fig.~\ref{fg:phase_diag}(d).  A change in sign from the normal state due to superconducting fluctuations has been predicted theoretically.\cite{ulla91,nish97}

\subsection{Magnetic susceptibility}
\label{sc:chi}

\begin{figure}[b]
\centerline{\includegraphics[width=3.3in]{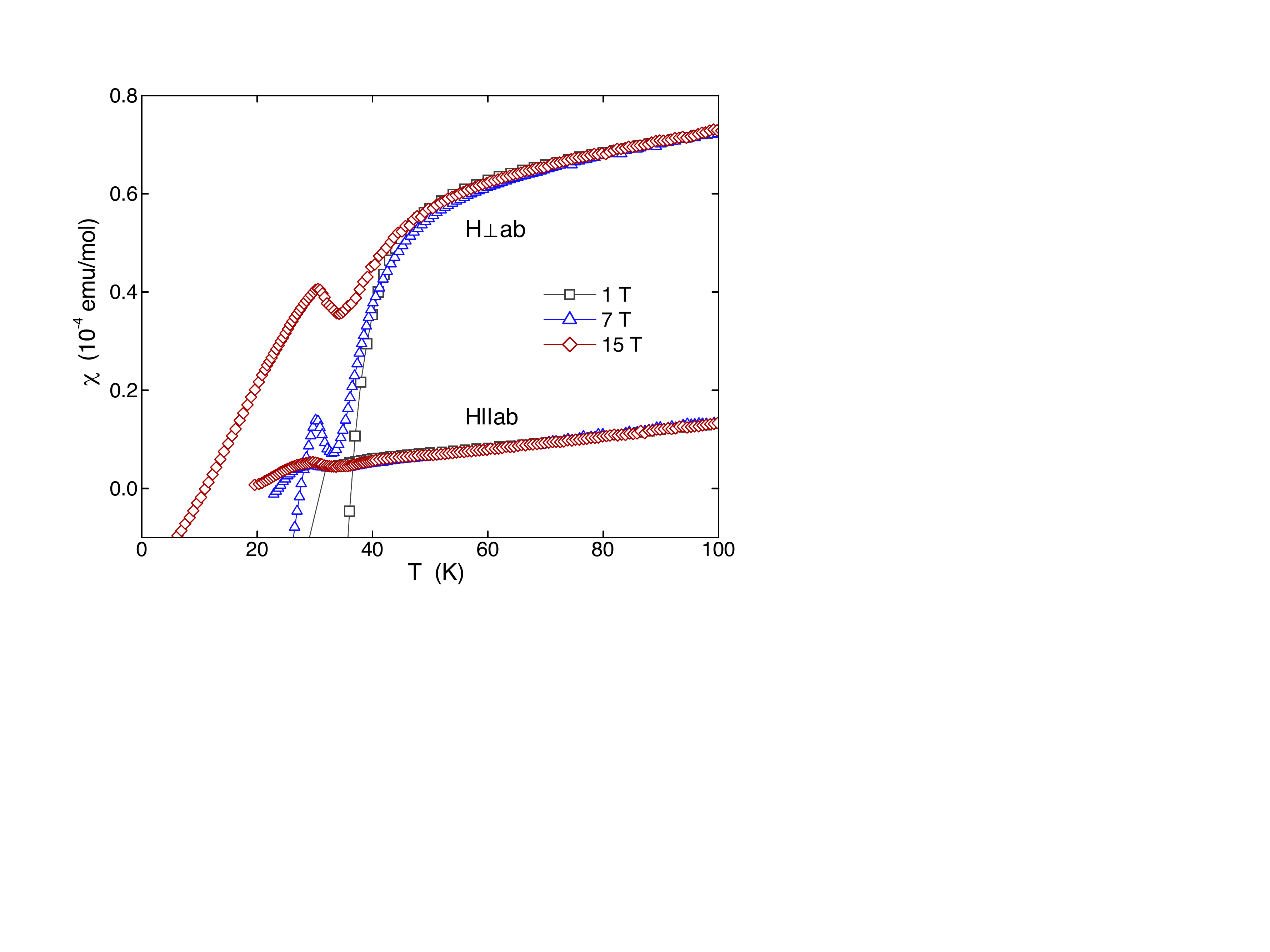}}
\caption{(color online) Magnetic susceptibility data measured for fields applied parallel and perpendicular to the $ab$ planes.  Data for $\mu_0H=7$ and 15 T were obtained with a VSM; the 1-T data are from Ref.~\onlinecite{huck08}.}
\label{fg:chi} 
\end{figure}

To confirm our analysis of superconducting contributions to $R_{\rm H}$, we present in Fig.~\ref{fg:chi} measurements of the magnetic susceptibility in fields up to 15~T.  One can see from $\chi(H\|ab)$, measured with fields parallel to the planes, that the normal state susceptibility decreases roughly linearly with temperature due to the paramagnetic response of Cu spins.\cite{huck08}  In contrast, there is a growing diamagnetic drop in $\chi(H\bot ab)$ as one cools, especially below $\sim40$~K.  The kink between 30 and 34~K is associated with a structural transition, discussed in Ref.~\onlinecite{wen12a}, that enables the appearance of weak stripe order, even in zero field. 

\begin{figure}[t]
\centerline{\includegraphics[width=3.3in]{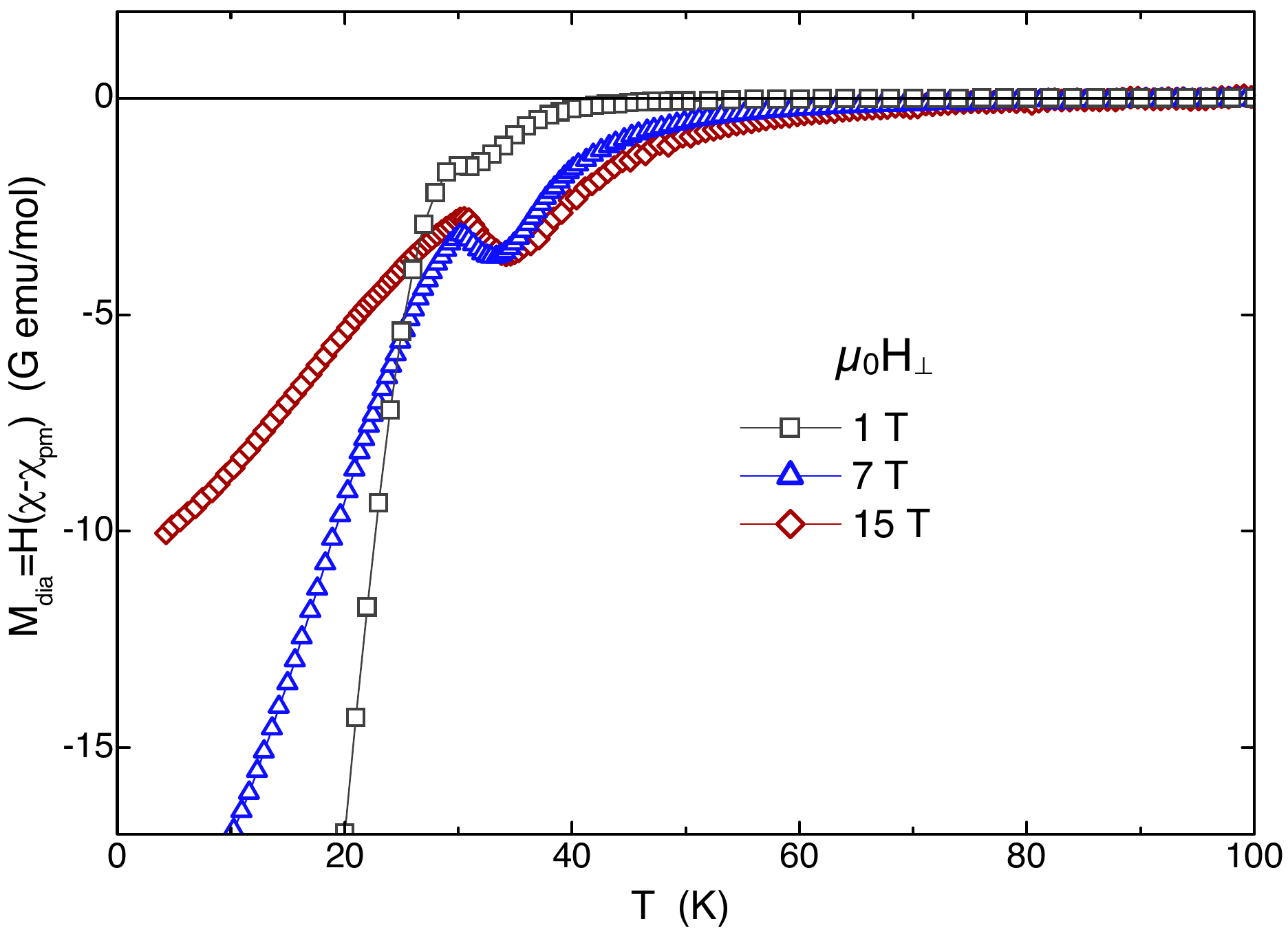}}
\caption{(color online) Diamagnetic magnetization obtained from $\chi(H\bot ab)$ data of Fig.~\ref{fg:chi} after subtracting a linear fit to the data between 80 and 100~K.  The high-field crossover at $T\sim30$~K is consistent with the appearance of the LVL state, as discussed in the text. }
\label{fg:Mdia} 
\end{figure}

To extract the diamagnetic response, a linear fit to $\chi(H\bot ab)$ between 80 and 100~K, representing the paramagnetic contribution $\chi_{\rm pm}$, has been extrapolated and subtracted from the data.  Multiplying by the field, we obtain the diamagnetic magnetization, $M_{\rm dia}$, that is plotted versus temperature for several $H_\bot$ in Fig.~\ref{fg:Mdia}.  We first note that the structural transition has a modest impact on the thermal evolution of the diamagnetism,\footnote{In a previous analysis, the effective change in the Josephson coupling caused by the transition\cite{home12} was invoked to explain the evolution of the diamagnetism measured with a small field.\cite{wen12a} For $\mu_0H_\bot=7$ and 15~T, however, there is no Josephson coherence between the layers, as indicated in Fig.~\ref{fg:phase_diag}(a), so that the oscillation in the susceptibility must be due to effects within the planes, possibly associated with the onset of weak stripe order.\cite{wen12}} which continues to grow on cooling below 30~K.    More significantly, one can see that $-M_{\rm dia}$ grows with field for $T\gtrsim35$~K but decreases with field for $T\lesssim25$~K.  Such behavior is qualitatively consistent with the predicted\cite{ogan06} response of a stack of decoupled superconducting layers with $T_{\rm BKT}\sim30~K$.  The observed response is also similar to that measured in magnetization studies of \lsco\ with $x=0.09$ (Ref.~\onlinecite{li10}) and $x=0.10$ (Ref.~\onlinecite{huh01}).

\subsection{Interlayer resistivity, $\rho_\bot(H_\bot,T)$}
\label{sc:perp}

Let us now consider $\rho_\bot(H_\bot)$ in the superconducting regime, as illustrated back in Fig.~\ref{fg:raw}(a).  For $T\lesssim30$~K, increasing $H_\bot$ initially causes $\rho_\bot$ to become finite, followed by rapid growth and eventual saturation, followed by a gradual decrease beyond the maximum.  Such behavior has been studied previously, especially in \bscco\ \cite{moro00,shib01} and Bi$_2$Sr$_{2-x}$La$_x$CuO$_{6+\delta}$.\cite{ono04}  The rise of $\rho_\bot$ with increasing $H_\bot$ is due to suppression of the conduction channel associated with interlayer pair tunneling; on crossing the maximum, single-particle transport dominates.\cite{moro00}  The region of negative magnetoresistance at high field has been attributed to the impact of $H_\bot$ on the pseudogap \cite{shib01}; reducing the antinodal gap increases the density of normal carriers that can move between planes.  Parallels have also been drawn with the field-tuned superconductor-insulator transition observed in disordered thin-films of various metals.\cite{stei05,aubi06,nguy09}  By this latter analogy, the resistive transition in $\rho_\bot$ can be viewed as a transition to a Cooper pair insulator phase at high $\mu_0H_\bot$.  In our case, the Cooper pairs are localized along the $c$ axis, becoming restricted to the CuO$_2$ layers.

\begin{figure}[b]
\centerline{\includegraphics[width=3.3in]{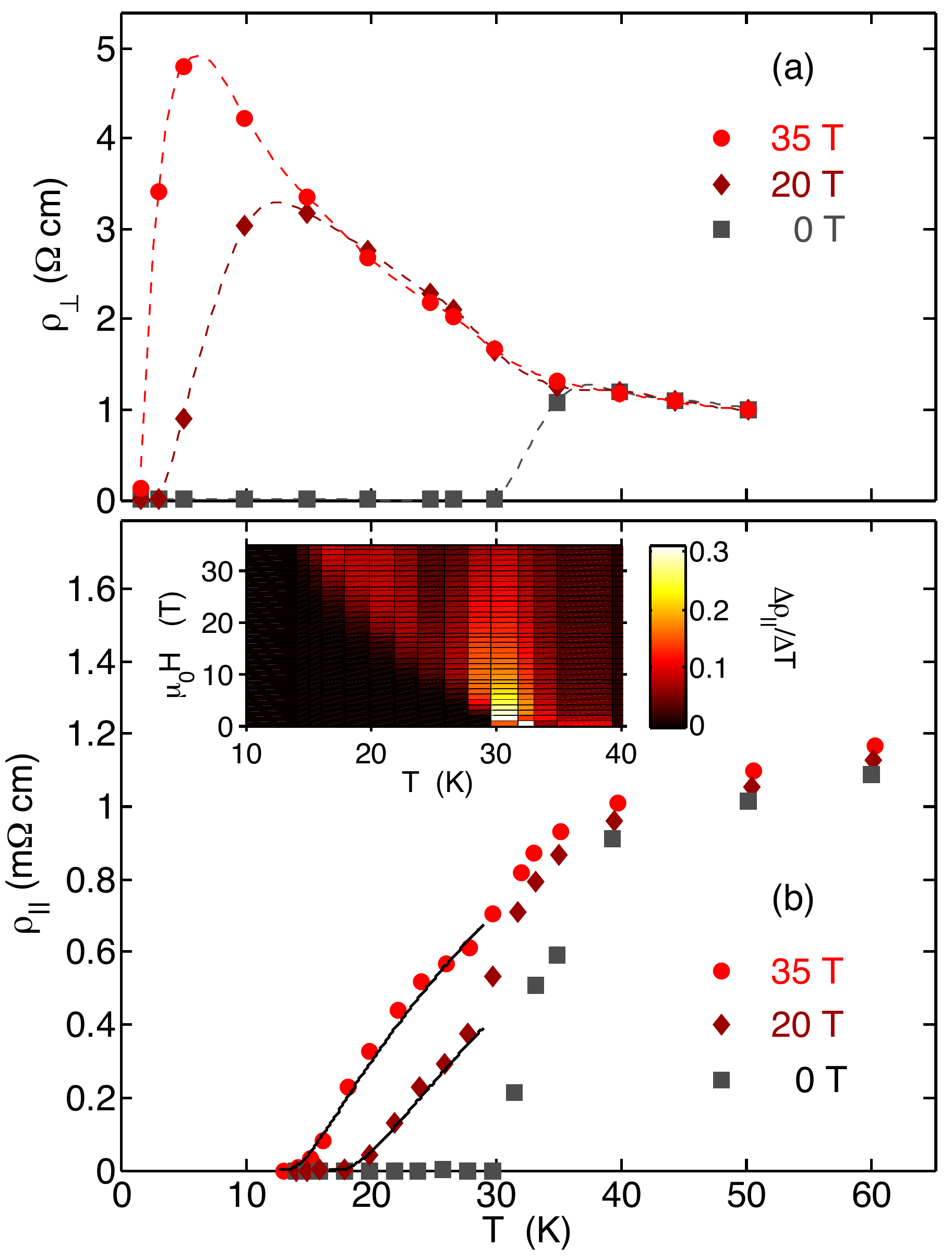}}
\caption{(color online) Measurements of (a) $\rho_\bot$ and (b) $\rho_\|$ as a function of temperature for $\mu_0H_\bot=0$, 20, and 35~T. The dashed lines in (a) are guides to the eye.  The solid lines in (b) are calculations of 2D flux-flow resistivity, as discussed in the text.  Inset of (b) shows $\Delta\rho_\|/\Delta T$ as a function of $H_\bot$ and $T$. }
\label{fg:tdep} 
\end{figure}

To emphasize the striking difference between $\rho_\bot$ and $\rho_\|$ in an applied field, we compare their temperature dependences in Fig.~\ref{fg:tdep}  for $\mu_0H_\bot=0$, 20, and 35~T.  For $\rho_\bot$, the field appears to shift the superconducting transition to low temperature.  In contrast, $\rho_\|$ shows a substantial drop near 30~K even in the highest field, and it continues towards zero on further cooling.   There is clearly a broad regime in which the superconducting layers are decoupled in terms of coherent Cooper-pair transport.

Now we want to be a bit more quantitative in defining transitions and crossovers.  The regime of 3D superconductivity ends when $\rho_\bot$ becomes finite.  We label the field at which this occurs as $H_c^\bot$.    Our determination of $H_c^\bot$ is indicated by the triangles in Fig.~\ref{fg:phase_diag}(a). 

To analyze the growth of $\rho_\bot$ with field, we start with the model of a stack of Josephson junctions between superconducting CuO$_2$ layers.\cite{lawr71}  It has been argued by several groups that the field-induced rise in $\rho_\bot$ can be understood in terms of phase fluctuations in the interlayer Josephson junctions due to thermal noise.\cite{bric91,gray93,kado94,hett95}  In this interpretation, the relevant quantity is the extensive resistance per Josephson junction.  Hettinger {\it et al.} \cite{hett95} demonstrated empirically that the effective junction area corresponds to $A = \Phi_0/(B_\bot+B_0)$, where $\Phi_0$ is the flux quantum and $B_0$ is a parameter.  Some of us have shown previously\cite{wen12} that this approach gives a good description of the evolution of $\rho_\bot(T,H_\bot)$ in our sample for $T<T_{c0}$ with $B_\bot\approx\mu_0H_\bot$ and $B_0=2.2$~T. The effective junction resistance is then $R_\bot = \rho_\bot s/A$, where $s$ is the interlayer spacing (6.6~\AA).   

According to Halperin {\it et al.}~\cite{halp10}, the criterion for a Josephson junction to become effectively insulating is that it exceed $R_Q=h/(4e^2)=6.45$~k$\Omega$, the quantum of resistance for Cooper pairs.   We define $H_Q$ as the field at which $R_\bot=R_Q$; the $T$ dependence of $H_Q$ is shown by the squares in Fig.~\ref{fg:phase_diag}(a).  As one can see, the separation between $H_Q$ and $H_c^\bot$ is rather modest.

\begin{figure}[t]
\centerline{\includegraphics[width=2.9in]{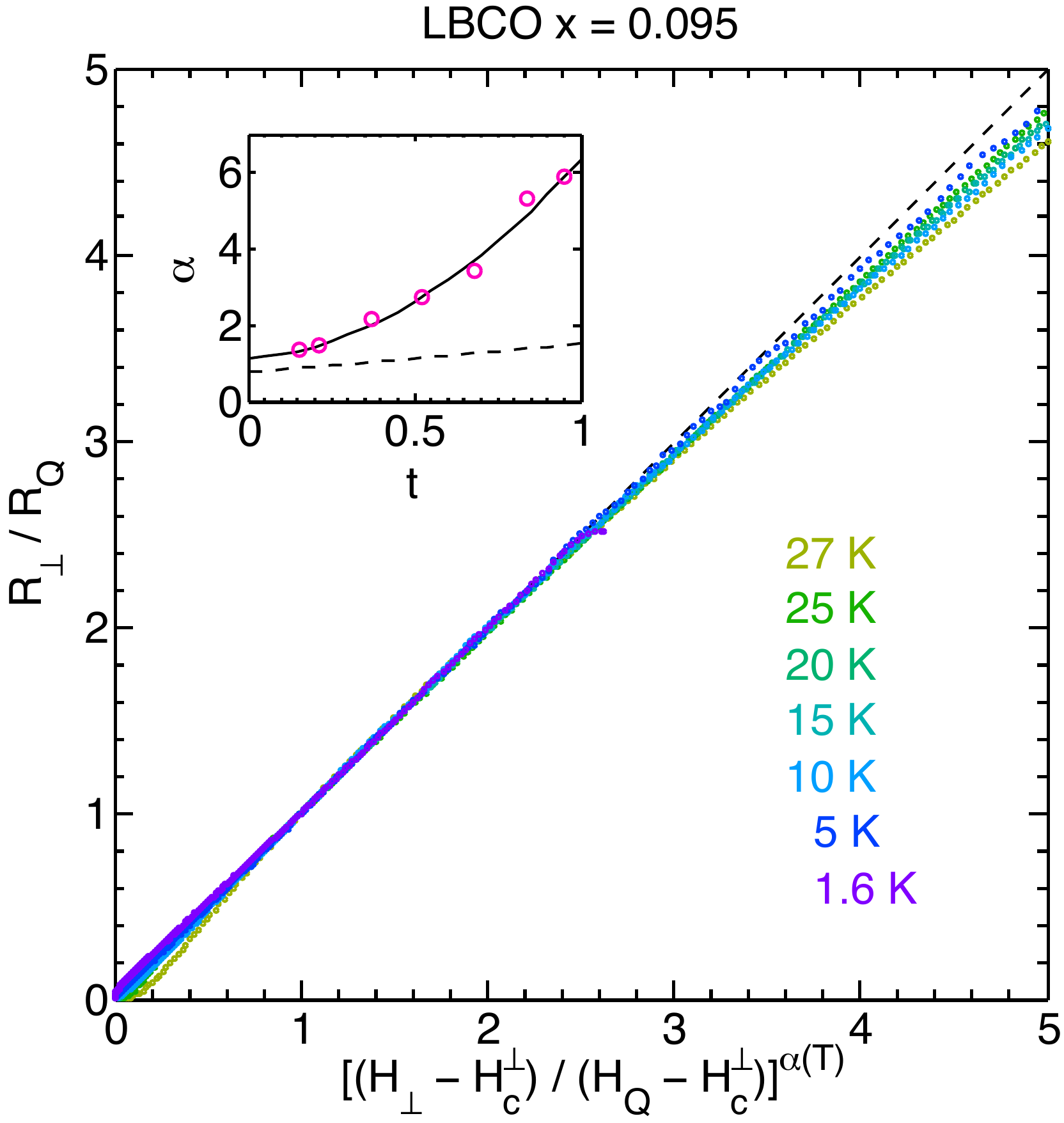}}
\caption{(color online) Plot of $R_\bot/R_Q$ vs.\ $[(H_\bot-H_c^\bot)/(H_Q-H_c^\bot)]^{\alpha(T)}$, as described in the text, for data at temperatures from 1.6 to 27~K.  The dashed straight line is for reference.  The $T$ dependence of the exponent $\alpha$ is plotted in the inset.  Dashed line represents $\alpha_{\rm K}$, as described in the text; solid line corresponds to $\alpha=\alpha_{\rm K}+0.35+4.5t^2$, with $t=(T_{c0}-T)/T_{c0}$ and $T_{c0}=31.5$~K.  }
\label{fg:RQ} 
\end{figure}

Looking at Fig.~\ref{fg:raw}(a), it appears that there is a common shape to $\rho_\bot(H_\bot)$ measured at different temperatures.  In Fig.~\ref{fg:RQ}, we show that $R_\bot/R_Q$ scales as $[(H_\bot-H_c^\bot)/(H_Q-H_c^\bot)]^{\alpha(T)}$, with the $T$ dependence of the exponent $\alpha$ displayed in the inset.  The scaling is motivated by a calculation from Konik \cite{koni11} for $\rho_\bot$ in a model of weak Josephson coupling between 2D layers; he predicts $\alpha_{\rm K}=\frac34(1+t)$, with $t=(T_{c0}-T)/T_{c0}$, which is represented by the dashed line in the inset.   

\subsection{In-plane resistivity, $\rho_\|(H_\bot,T)$}
\label{sc:par}

We have already noted that the Hall effect and susceptibility measurements indicate the onset of strong superconducting fluctuations below 40~K, in a fashion that is surprisingly independent of field.  We see related behavior of $\rho_\|$ in Fig.~\ref{fg:tdep}(b), where the high-field data show a rapid drop at $\sim30$~K.  To emphasize this behavior, the inset of Fig.~\ref{fg:tdep}(b) shows the ratio of finite differences $\Delta\rho_\|/\Delta T$, as a function of field and temperature.  
For $\mu_0H_\bot\gtrsim 1$~T, one can see that the maximum of this approximate derivative occurs at $30\pm2$~K, which we identify as $T_{c}^{\rm 2D}$, the onset of the LVL state.\footnote{We note that there is also structure in $\rho_{\parallel}(T)$ for zero field and $T>T_{c}$; however, that is associated with 3D superconducting fluctuations as discussed in \protect\cite{wen12a}.}  The finite resistivity at lower temperature indicates a lack of phase order, as we discuss next.  

BKT predicted that, in a 2D system with vortex-like excitations, it is possible to have a topological transition from an ordered to a phase-disordered state.\cite{bere71,kost73}  The transition can be described as an unbinding of thermally-excited vortex-antivortex pairs.  The nature of the transition depends crucially on having the interaction energy of a pair of vortices vary logarithmically as their separation distance.  For a superconductor, the logarithmic interaction applies only at distances shorter than the magnetic penetration depth; at larger distances it is screened.\cite{kost73}  In a thin film, it is possible to enhance the effective screening length,\cite{pear64,beas79} but this can still be smaller than the sample size.  Attempts to observe BKT transitions in thin films have been controversial,\cite{gray85,gasp12} and there have been analyses showing that effects near the edges of a thin film could give the appearance of BKT behavior even when it is absent in the bulk of the film.\cite{gure08}

This history would make it appear that consideration of BKT-like effects in our bulk sample would be inappropriate.  It turns out, however, that the presence of many adjacent, phase-decoupled layers restores the conditions necessary for a BKT-like transition.  Raman, Oganesyan, and Sondhi\cite{rama09} have shown that, due to the electromagnetic interactions of pancake vortices\cite{clem91} in different layers, the interaction energy between vortices remains logarithmic to long distances.  They find that the system does exhibit a phase disordering transition, though there are small quantitative corrections relative to the predictions of the 2D theory.\cite{rama09}  This analysis provides an explanation for the BKT-like transition observed in LBCO $x=1/8$ in zero field.\cite{li07}

\begin{figure}[t]
\centerline{\includegraphics[width=3.3in]{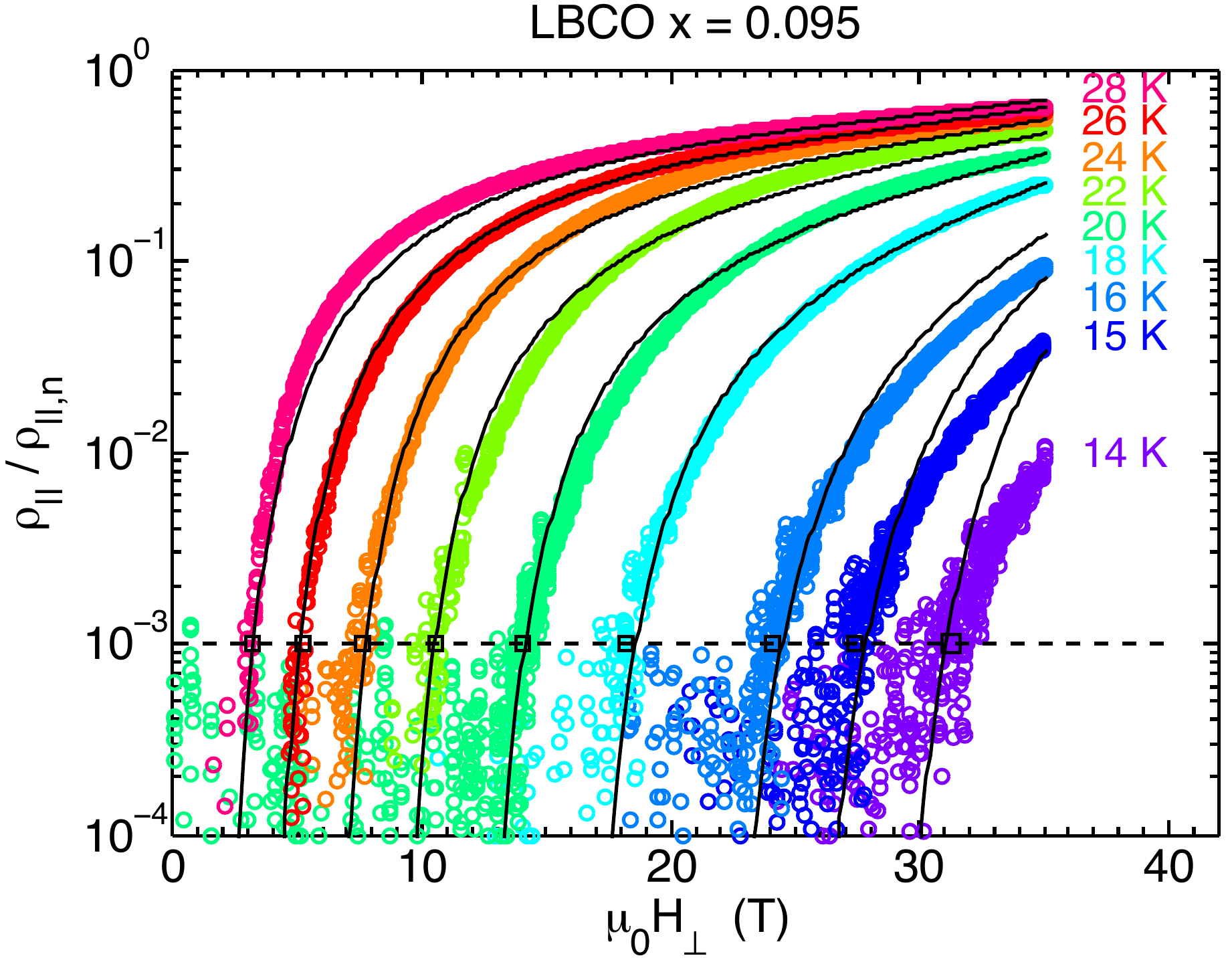}}
\caption{(color online) Plot of $\rho_\|(H_\bot)$ normalized to approximate normal-state values (see text).  Squares indicate fields at which data cross $10^{-3}$ (dashed line).  Solid lines are calculations using Eq.~(\ref{eq:kost}), as discussed in the text. }
\label{fg:rho_2D} 
\end{figure}

With that context, let us consider Fig.~\ref{fg:rho_2D}, where we plot $\rho_\|(H_\bot)$, normalized to the normal-state $\rho_n(H_\bot)$ evaluated in Sec.~\ref{sc:normal}, for a number of temperatures.  At each temperature, there is a phase-transition-like rise in the normalized resistivity as $H_\bot$ increases.  To make an initial estimate of the transition field, which we label $H_c^{\|i}$, we choose the point at which the $\rho_\|/\rho_n$ reaches $10^{-3}$, as indicated by the squares superimposed on the data in Fig.~\ref{fg:rho_2D}.  The obtained transition fields correspond approximately to the circles plotted in the phase diagram Fig.~\ref{fg:phase_diag}(b). (The corrected values of $H_c^\|$ are discussed below.)  Comparing with Fig.~\ref{fg:phase_diag}(a), we see that $H_c^\| > H_Q$ for any $T$, so the apparent transition occurs in the regime where there is no coherent Josephson tunneling between layers.   Thus, on cooling in a field of $\mu_0H_\bot\gtrsim1$~T, we see behavior consistent with a BKT-like transition from the LVL state to the LPD-SC state.

To go further, we test the functional form of critical behavior of the resistivity.  Above a BKT transition at $T_{\rm BKT}$, the resistivity, which is proportional to the product of the density of free vortices and the vortex mobility,  is predicted\cite{kost74,halp79} to have the form
\begin{equation}
  \rho_\| / \rho_{\|,n} = a e^{-b/\tau^\gamma},
  \label{eq:kost}
\end{equation}
where $\tau=(T-T_{\rm BKT})/T_{\rm BKT}$, $a$ and $b$ are constants of order one, and $\gamma=0.5$.   In the mixed state, the applied $H_\bot$ will increase the density of free vortices, and hence will increase the resistivity.\cite{doni79,minn81}  In a study of 2D-like superconductivity in LBCO $x=1/8$,\cite{li07} it was found empirically that $\rho_\|(T)$ in a field could still be described by Eq.~(\ref{eq:kost}), provided that one takes account of the reduction of \tbkt\ by the field.  Theoretically, one does not expect \tbkt\ to remain finite in the mixed state of a 2D superconductor \cite{fish91}; however, our system is never truly 2D.  Perhaps the electromagnetic interactions between pancake vortices in neighboring layers, considered by Raman {\it et al.}\cite{rama09} in the zero-field limit, are sufficient to maintain a finite $T_{\rm BKT}$ in large $H_\bot$.    

Without theoretical justification, we take Eq.~(\ref{eq:kost}) as a useful functional form.   We use it to model the field-dependent data of Fig.~\ref{fg:rho_2D} by inserting the field dependence through $T_{\rm BKT}(H_\bot)$; the latter corresponds to $H_c^\|(T)$, which we have already estimated.  Of course, our estimates $H_c^{\|i}$ were determined at $\rho_\|/\rho_{\|,n}=10^{-3}$; we need to adjust these values for the finite cutoff.  Empirically, we find that an effective one-parameter formula for the correction is $\mu_0H_c^\| = \mu_0H_c^{\|i}-C_i/T$ with $C_i=44.8$\,T\,K.  Each $H_c^\|$ value corresponds to a particular temperature, which can be viewed as $T_{\rm BKT}$ for that field value.  We fit $T_{\rm BKT}(H_\bot)$ with a cubic polynomial in $H_\bot$.  Plugging these values into Eq.~(\ref{eq:kost}), we obtain the curves indicated by the solid (black) lines in Fig.~\ref{fg:rho_2D}, using the fixed set of parameters ($a=2.5$, $b=1.5$, $\gamma=0.6$).  We see that the data are reasonably well described simply by accounting for the variation in transition temperature with field.

We can use the same formula and parameters to describe the temperature dependence of the resistivity in the LVL phase at fixed field.  The solid lines through the data points in Fig.~\ref{fg:tdep}(b) correspond to such calculations, with $T_{\rm BKT}=16.5$ and 12.5~K for $\mu_0H_\bot=20$ and 35~T, respectively.  

\section{Summary and discussion}
\label{sc:summ}

We have seen that decoupling the CuO$_2$ planes with a transverse magnetic field reveals a crossover at $\sim30$~K to a layered vortex liquid state, with the crossover being surprisingly insensitive to the strength of the field.  The development of the strong superconducting correlations within the layers is evident in $\rho_\|$, $R_{\rm H}$, and the anisotropic magnetic susceptibility.  With further cooling, there is an apparent transition to a layered, phase-decoupled superconducting state.  This is detected through a transition-like drop in $\rho_\|$; the phase decoupling is clear from the behavior of $\rho_\bot$.

A vortex-liquid state has previously been proposed to explain features of the pseudogap phase at $T>T_{c0}$ \cite{ande07c,tesa08}.  Along with other recent work \cite{bilb11b,mosq11},  our results suggest that such a scenario is overly optimistic.  The rise of $\rho_\|$ on warming through $T_c^{\rm 2D}$ indicates a loss of uniform superconducting correlations in the normal state.   $T_{c0}$ is  slightly larger than $T_c^{\rm 2D}$, suggesting that, upon cooling in zero field, 3D order develops just before the individual layers would become superconducting in the absence of Josephson coupling, as others have observed in \bscco\ \cite{cors99,wan96,weye09}.

\begin{figure}[t]
\centerline{\includegraphics[width=3.3in]{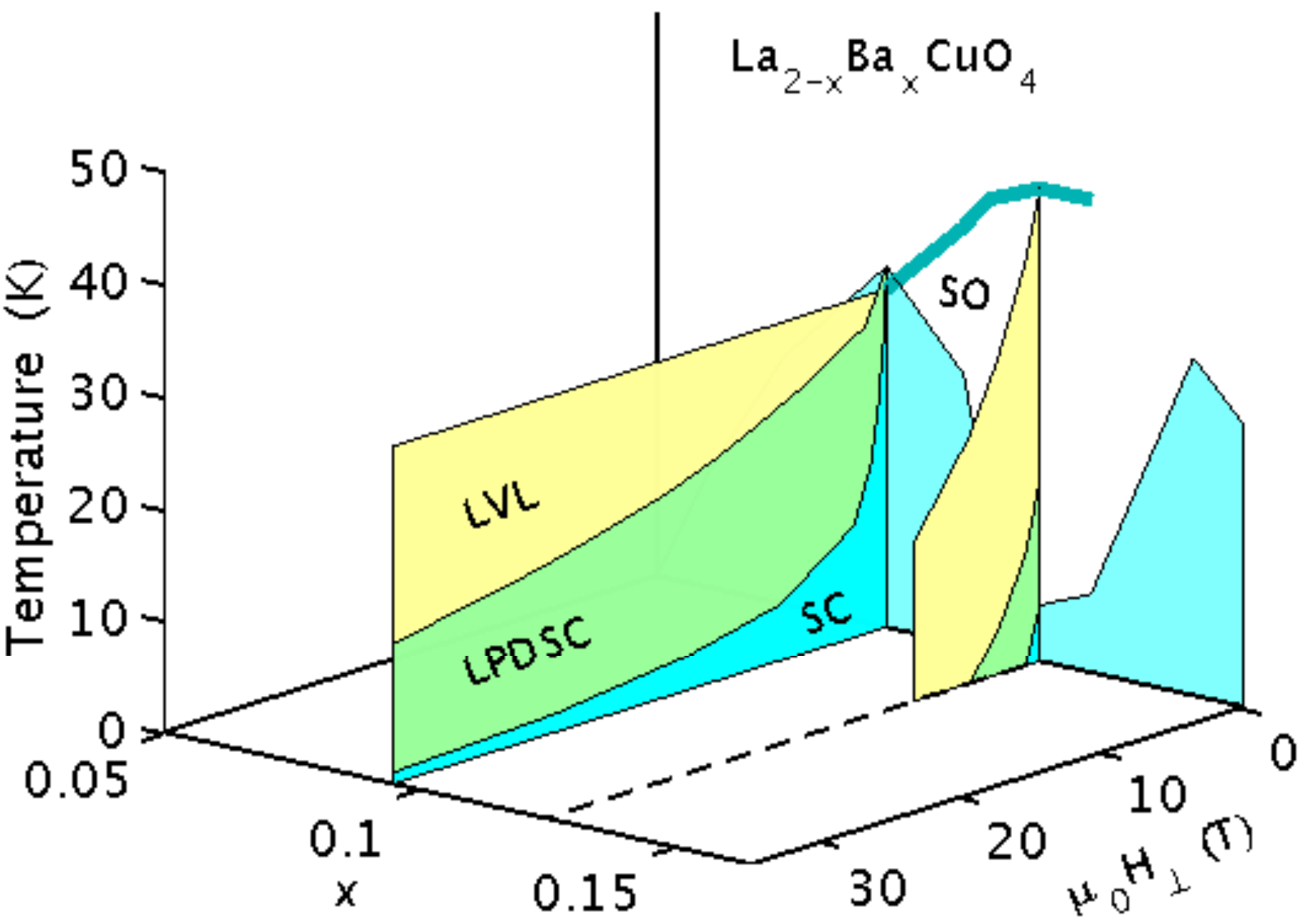}}
\caption{(color online) Phase diagram for LBCO as a function of $T$, $H_\bot$, and $x$ comparing results for various samples.   For the $T$-$x$ plane at $H_\bot = 0$, spin order (SO) sets in below the thick line, and superconductivity (SC) occurs in the shaded region below the thin line.\cite{huck11} The field-dependent results for $x=0.095$ are from Fig.~\ref{fg:phase_diag}, and the results for $x=1/8$ are from Ref.~\onlinecite{li07}. }
\label{fg:htx} 
\end{figure}

The LVL state that we observe here only at finite $H_\bot$ is equivalent to the state previously detected in LBCO $x=1/8$ below 40~K in zero field.\cite{li07}  We compare the phase diagrams of these two compositions in Fig.~\ref{fg:htx}.  In both cases, the LPD-SC state is reached at lower temperature.  The observation of such a state is only possible when the interlayer Josephson coupling is frustrated.  We note that one proposed origin of the frustration is the development of pair-density-wave (PDW) order in association with charge and spin stripe order \cite{hime02,berg07,berg09b}.  Neutron and x-ray diffraction measurements on LBCO $x=0.095$ have demonstrated that both charge and spin stripe order are enhanced by $H_\bot$ for $T\lesssim T_c^{\rm 2D}$ (Refs.~\onlinecite{wen12,huck13}); however, it should be noted that, while the occurrence of PDW order would explain the loss of 3D superconducting order, it would not, by itself, explain the apparent stability of the LPD-SC state in large $H_\bot$.

The modulated pair wave function of the PDW state provides a way for the superconductivity to coexist with local antiferromagnetic order: the superconducting wave function has zeros at the extrema of the spin density wave, and vice versa.  This is consistent with experimental evidence that long-range commensurate antiferromagnetic order and superconductivity do not coexist in LBCO or \lsco.\cite{birg06,bozo03,sute11}  At the same time, the modulation makes the PDW state quite sensitive to disorder, consistent with the strongly depressed bulk $T_c$ in LBCO $x=1/8$.  For $x=0.095$, the superconducting order develops at much higher $T_{c0}$, and $T_c^{2D}$ is virtually independent of $H_\bot$ even for $\mu_0H_\bot$ as high as 35 T.  The stability indicates that there are at least quantitative differences from $x=1/8$.  

One possible interpretation of the field-independence of $T_c^{\rm 2D}$ is that this crossover is determined by competition between different correlations.  For example, Emery, Kivelson, and Zachar\cite{emer97} originally proposed that superconductivity in a stripe-ordered system would involve in-phase Josephson coupling between neighboring charge stripes.  In contrast, the PDW state is proposed to have antiphase coupling.\cite{hime02,berg07}  Before relative phase order is established, there may be a competition between the interactions that favor in-phase vs.\ anti-phase coupling.  In this scenario, strong pairing correlations would exist within the fluctuating charge stripes at $T>T_c^{\rm 2D}$, but they would have only a weak impact on measurable quantities.  The onset of interstripe phase coherence might be relatively insensitive to $H_\bot$.  Theoretical analysis is necessary to determine whether this speculation is realistic.  We note that there is an empirical correlation between the onset of spin-stripe order and $T_{c0}$ in several cuprates,\cite{li07,lake02,lee04a} with the onset of spin-stripe order exhibiting minimal dependence on $H_\bot$.\cite{lake02,lee04a,wen08b}  This suggests that the correlations within spin stripes can impact the development of superconducting phase order, in samples with varying types of superconducting correlations.

We noted in Sec.~\ref{sc:perp} that the the temperature dependence of $\rho_\bot$ in large $H_\bot$ displayed in Fig.~\ref{fg:tdep}, with a rapid rise and gradual fall off with increasing $T$, has been seen previously in other cuprates.\cite{moro00,shib01,ono04}  There have also been studies of $\rho_\bot$ in underdoped \ybco\ for transverse fields up to 60 T by Vignolle {\it et al.}\cite{vign12}  They have been able to measure into the regime where $\rho_\bot$ remains finite and large down to 2~K.  Assuming that they have reached the normal state, they make a correction for normal-state magnetoresistance and obtain results suggesting a crossover to coherent $c$-axis conduction at low temperature.  While we have not measured to such high fields, we nevertheless have observed a regime in which $\rho_\bot$ exhibits an insulator-like temperature dependence while superconducting correlations are present within the planes.  When there is a metallic-like temperature dependence of $\rho_\bot$, it is due to superconductivity within the layers.  While Vignolle {\it et al.}\cite{vign12} have provided self-consistent arguments to support their identification of normal state behavior at high-fields in \ybco, we suggest that the possibility of hidden superconducting correlations within the CuO$_2$ bilayers that begin to impact $\rho_\bot$ at sufficiently low temperature should be considered as a possible alternative explanation.

Another intriguing observation in \ybco\ at high field and low temperature is the negative value of $R_{\rm H}$, with a magnitude even larger than at $T>T_c$.\cite{vign12,lebo11}  This behavior has been interpreted as evidence for electron pockets associated with the high-field normal state.\cite{lebo07}  While we have also observed a regime of negative $R_{\rm H}$ in our LBCO $x=0.095$ sample, the magnitude of $R_{\rm H}$ is much smaller.  The correspondence of this regime with the LPD-SC state, as indicated in Fig.~\ref{fg:phase_diag}(d), suggests in our case that it may be associated with superconducting fluctuations.\cite{ulla91,nish97}

Ramshaw {\it et al.}\cite{rams12} have recently made the interesting observation for \ybco\ that, at low temperature, the magnetic field at which $\rho_\bot$ becomes finite, which they label $H_{c2}$, is a minimum for a hole concentration of $\sim0.12$.  It appears that LBCO also follows this pattern, and it will be interesting to see whether other cuprate families follow it.   In terms of notation, the field at which the resistivity becomes finite is more commonly labelled as the irreversibility field, $H_{\rm irr}$; $H_{c2}$ should correspond to the field at which pairing is completely eliminated.  For underdoped LBCO, at least, our results suggest that $H_{c2}$ cannot be readily determined from measurements of $\rho_\bot$.  Even when measurements sensitive to in-plane correlations are made, it appears that $H_{c2}$ for LBCO $x=0.095$ is quite large for temperatures all the way to $T_{c0}$.

\acknowledgments

We thank E. Fradkin, S. A. Kivelson, R. M. Konik, and A. Tsvelik for stimulating discussions and valuable guidance.  The efforts of SJH, JW, GG, QL, and JMT on the magneto-transport work were supported by the Center for Emergent Superconductivity, an Energy Frontier Research Center funded by the U.S. Department of Energy's Office of Basic Energy Sciences (BES), Division of Materials Science and Engineering (DMSE).  The magnetic susceptibility measurements by MH were supported  through Brookhaven by the U.S. DOE, Office of BES, DMSE under Contract No.\ DE-AC02-98CH10886.  The National High Magnetic Field Laboratory is supported by the State of Florida and the National Science Foundation's Division of Materials Research through DMR-0654118.

\bibliography{lno,theory}

\end{document}